\documentclass[twocolumn,aip]{revtex4}

\usepackage{graphicx}
\usepackage{dcolumn}
\usepackage{bm}
\usepackage{color}
\usepackage{setspace}

\begin{document}


\title{Transverse thermoelectric effect in La$_{0.67}$Sr$_{0.33}$MnO$_{3}$$\mid$SrRuO$_{3}$ superlattices}

\author{Y. Shiomi$^{\, 1,2}$} 
\author{Y. Handa$^{\, 1}$}
\author{T. Kikkawa$^{\, 1}$}
\author{E. Saitoh$^{\, 1,2,3,4,5}$}
\affiliation{$^{1}$
Institute for Materials Research, Tohoku University, Sendai 980-8577, Japan }
\affiliation{$^{2}$
Spin Quantum Rectification Project, ERATO, Japan Science and Technology Agency, Aoba-ku, Sendai 980-8577, Japan
}
\affiliation{$^{3}$
WPI Advanced Institute for Materials Research, Tohoku University, Sendai 980-8577, Japan
}
\affiliation{$^{4}$
CREST, Japan Science and Technology Agency, Tokyo 102-0076, Japan
}
\affiliation{$^{5}$
Advanced Science Research Center, Japan Atomic Energy Agency, Tokai 319-1195, Japan
}

\date{\today}

\begin{abstract}
Transverse thermoelectric effects in response to an out-of-plane heat current have been studied in an external magnetic field for ferromagnetic superlattices consisting of La$_{0.67}$Sr$_{0.33}$MnO$_{3}$ and SrRuO$_{3}$ layers. The superlattices were fabricated on SrTiO$_{3}$ substrates by pulsed laser deposition. We found that the sign of the transverse thermoelectric voltage for the superlattices is opposite to that for La$_{0.67}$Sr$_{0.33}$MnO$_{3}$ and SrRuO$_{3}$ single layers at $200$ K, implying an important role of spin Seebeck effects inside the superlattices. At $10$ K, the magnetothermoelectric curves shift from the zero field due to an antiferromagnetic coupling between layers in the superlattices.  
\end{abstract}

\maketitle

Spin Seebeck effects (SSEs) \cite{uchida-nature} which enable electricity generation via spin currents as a result of a temperature gradient is a promising candidate for thermoelectric applications \cite{kirihara}. Along with the success in the spin caloritronics which focuses on the interaction of spins with heat currents \cite{bauer, heremans-review} and the discovery of the SSE \cite{uchida-nature}, Nernst-Ettingshausen (Nernst) effects in ferromagnetic conductors have also gained interest in the spintronics field. Nernst effects are the thermoelectric counterparts of Hall effects, {\it viz.} generation of a transverse electric field by a longitudinal thermal gradient in the presence of an external magnetic field. 
In ferromagnetic conductors, anomalous Nernst effects (ANEs), which are caused by the spin-orbit interaction and proportional to the magnetization curve, also appear. ANEs in magnetic conductors have intensively been studied in the field of condensed-matter physics in terms of the topological nature for Bloch electrons \cite{ong, xiao, chiba, onoda, hanasaki, shiomi-Nernst}. 
\par

Whereas spintronics or spin-caloritronics experiments have commonly been carried out using conventional alloys and ferrites, perovskite-type oxides have received attention from the wide science community because of a rich variety of electronic properties \cite{tokura}. Especially, heterostructures of these perovskite oxide materials provide a fertile ground for novel physical phenomena related with interfaces \cite{hwang}. In the Seebeck effect, for example, a positive Seebeck coefficient was observed for superlattices made of YBa$_{2}$Cu$_{3}$O$_{7-\delta}$ and La$_{2/3}$Ca$_{1/3}$MnO$_{3}$, in spite of the negative values for a simple YBa$_{2}$Cu$_{3}$O$_{7-\delta}$ film and a La$_{2/3}$Ca$_{1/3}$MnO$_{3}$ film each \cite{heinze}. The sign change in thermopower by forming superstructures was attributed to an interface effect \cite{heinze}, which suggests that perovskite-based superlattices are an attracting stage for novel spin caloritronic effects.  
\par

A ferromagnetic superlattice comprising La$_{0.67}$Sr$_{0.33}$MnO$_{3}$ and SrRuO$_{3}$ is the target material in the present study. Both La$_{0.67}$Sr$_{0.33}$MnO$_{3}$ and SrRuO$_{3}$ are known to be ferromagnetic metals; La$_{0.67}$Sr$_{0.33}$MnO$_{3}$ is a soft magnet with high Curie temperature $T_{C} \approx 350$ K, while SrRuO$_{3}$ is a hard magnet with low $T_{C} \approx 150$ K. A novel magnetic property in this superlattice is an interfacial antiferromagnetic coupling between La$_{0.67}$Sr$_{0.33}$MnO$_{3}$ and SrRuO$_{3}$ layers, which originates from the hybridization of $2p$ state of O atoms with $3d$ states of Mn atoms and $4d$ states of Ru atoms \cite{ziese, lee}. The interlayer antiferromagnetic coupling induces an exchange bias effect \cite{ke}; the magnetization loop is shifted so that it is no longer symmetric about the zero magnetic field, as commonly implemented in ferromagnet$\mid$antiferromagnet interfaces. Whereas the magnetic properties of La$_{0.67}$Sr$_{0.33}$MnO$_{3}$$\mid$SrRuO$_{3}$ superlattices have been investigated well so far \cite{ke, ziese, lee, qzhang, padhan, kim, ke2, zieseAPL, zieseReview}, there are few studies of cross-plane magnetotransport properties which should strongly reflect the interlayer magnetic coupling.    
\par

In the present letter, we study a magnetothermoelectric effect along the Hall direction driven by a heat current transmitting across La$_{0.67}$Sr$_{0.33}$MnO$_{3}$$\mid$SrRuO$_{3}$ ferromagnetic superlattices using a so-called longitudinal setup for the measurement of SSEs \cite{kuchida} [see also Fig. \ref{fig2}(c)]. High-quality La$_{0.67}$Sr$_{0.33}$MnO$_{3}$$\mid$SrRuO$_{3}$ superstructures which were fabricated by pulsed laser deposition show a strong interlayer antiferromagnetic coupling below $105$ K. The measured magnetothermoelectric voltage was found to have an opposite sign to that for a La$_{0.67}$Sr$_{0.33}$MnO$_{3}$ single layer film and a SrRuO$_{3}$ single layer film at $200$ K, which is ascribable to an electronic reconstruction nucleated at the interfaces or generation of spin-current induced voltage in addition to the ANE. At $10$ K, clear shifts of hysteresis loops of the transverse thermoelectric voltages were observed. The directions of shifts depend on field-cooling processes, consistent with the exchange bias effect.  
\par  

\begin{figure}[t]
\begin{center}
\includegraphics[width=8.5cm]{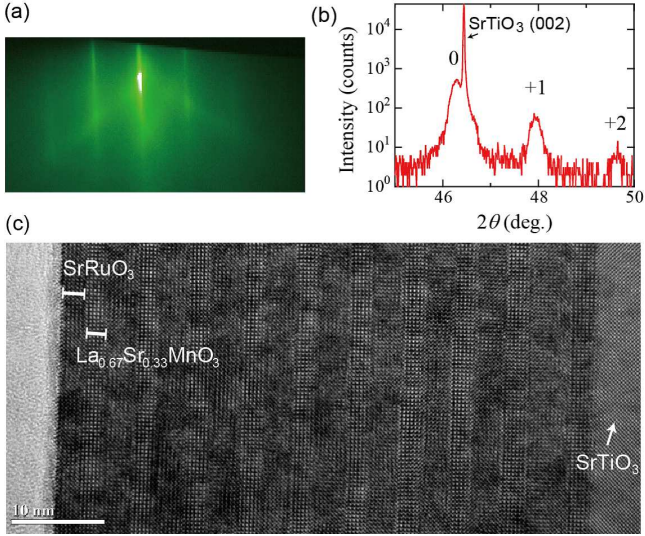}
\caption{(a) A streaky RHEED pattern of a La$_{0.67}$Sr$_{0.33}$MnO$_{3}$$\mid$SrRuO$_{3}$ superlattice sample. (b) X-ray diffraction pattern around the SrTiO$_{3}$ (002) reflection peak. Superlattice peaks are numbered $0$, $+1$, and $+2$. (c) An overview cross-sectional TEM image for a superlattice sample.} 
\label{fig1}
\end{center}
\end{figure}

Epitaxial films were grown on $0.5$-mm-thick SrTiO$_{3}$ (001) substrates by pulsed laser deposition from polycrystalline targets using a KrF excimer laser. Oxygen partial pressure was $0.3$ torr and substrate temperature was kept at $800$ $^{\circ}$C during the laser ablation. A La$_{0.67}$Sr$_{0.33}$MnO$_{3}$ layer was first grown on a SrTiO$_{3}$ substrate and then a SrRuO$_{3}$ layer was deposited. The above bilayer was repeated $10$ times. After deposition, samples were annealed at $800$ $^{\circ}$C in the $400$ torr oxygen atmosphere and then cooled to room temperature. The samples were characterized by Reflection High Energy Electron Diffraction (RHEED), X-ray diffraction, and transmission electron microscopy (TEM). We have confirmed that the grown samples show metallic resistivity below $300$ K; the sheet resistance at room temperature is about $150$ ${\rm \Omega}$ and residual-resistance ratio is $\sim 2$. Magnetization and magnetothermoelectric-effect measurements were performed in a Physical Property Measurement System (Quantum Design, Inc.), where an external magnetic field was applied parallel to the superlattices.
\par

We show a RHEED pattern taken along $[111]$ at an ambient temperature for a superlattice sample in Fig. \ref{fig1}(a). Clear streaks are observed, which indicate a flat surface of the grown film. Figure \ref{fig1}(b) shows a $\theta$-$2\theta$ X-ray diffraction scan for the same sample around the SrTiO$_{3}$ (002) reflection peak. Satellite peaks which support the superlattice structure are clearly observed, as numbered in Fig. \ref{fig1}(b). From the peak positions, the superlattice period is calculated using $\lambda/(2|\sin\theta_{i}-\sin\theta_{i+1}|)$, where $\lambda=0.154$ nm for Cu K${\alpha}$ radiation \cite{ziese2}. The obtained value is $5.7 \pm 0.2$ nm, which corresponds to the thickness of the La$_{0.67}$Sr$_{0.33}$MnO$_{3}$$\mid$SrRuO$_{3}$ bilayer.   
\par

\begin{figure}[t]
\begin{center}
\includegraphics[width=8.5cm]{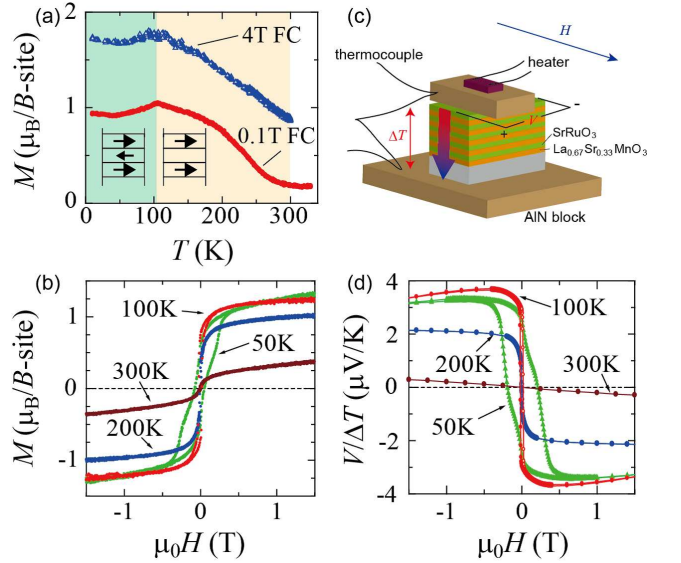}
\caption{(a) Temperature ($T$) dependence of the magnetization ($M$) measured under an in-plane magnetic field of $0.1$ T or of $4$ T for a La$_{0.67}$Sr$_{0.33}$MnO$_{3}$$\mid$SrRuO$_{3}$ superlattice. (b) Magnetic field ($H$) dependence of the magnetization ($M$) at some temperatures. The magnetic field is applied to an in-plane direction. (c) A schematic illustration of the measurement of thermoelectric voltage along the Hall direction. (d) Magnetic field ($H$) dependence of the transverse thermoelectric voltage divided by temperature difference ($V/\Delta T$) at some temperatures.    }
\label{fig2}
\end{center}
\end{figure}

Figure \ref{fig1}(c) shows a cross-sectional TEM image for a superlattice sample. The clear superlattice structure which has structural integrity of La$_{0.67}$Sr$_{0.33}$MnO$_{3}$ and SrRuO$_{3}$ layers was confirmed. Also, all the interfaces between La$_{0.67}$Sr$_{0.33}$MnO$_{3}$ and SrRuO$_{3}$ layers are sharp. The thicknesses of La$_{0.67}$Sr$_{0.33}$MnO$_{3}$ and SrRuO$_{3}$ layers were determined as $\sim 2.4$ nm and $\sim 3.5$ nm, respectively, and the total thickness of the superlattice was $57$ nm. These values are in agreement with those deduced from the X-ray diffraction measurements.      
\par

In Fig. \ref{fig2}(a), we show the temperature ($T$) dependence of magnetization, $M$, which was measured in the field-cooled condition under $\mu_{0}H=0.1$ T. Here, the diamagnetic contribution of the SrTiO$_{3}$ substrate was subtracted from the raw data. As $T$ decreases, a sharp increase in $M$ is observed below $\sim 300$ K, which corresponds to the ferromagnetic transition of the La$_{0.67}$Sr$_{0.33}$MnO$_{3}$ layers. The magnitude of $M$ increases with decreasing $T$, but below $105$ K, where the SrRuO$_{3}$ layers undergo a ferromagnetic transition, a clear decrease in $M$ is observed. This is evidence that the magnetization of the SrRuO$_{3}$ layers antiferromagnetically couples to that of the La$_{0.67}$Sr$_{0.33}$MnO$_{3}$ layers. The decrease in $M$ at $105$ K is observed also at $4$ T, demonstrating that the interlayer antiferromagnetic coupling is highly strong compared with similar superlattices reported in former studies \cite{zieseAPL, zieseReview}.
\par

\begin{figure}[t]
\begin{center}
\includegraphics[width=8.5cm]{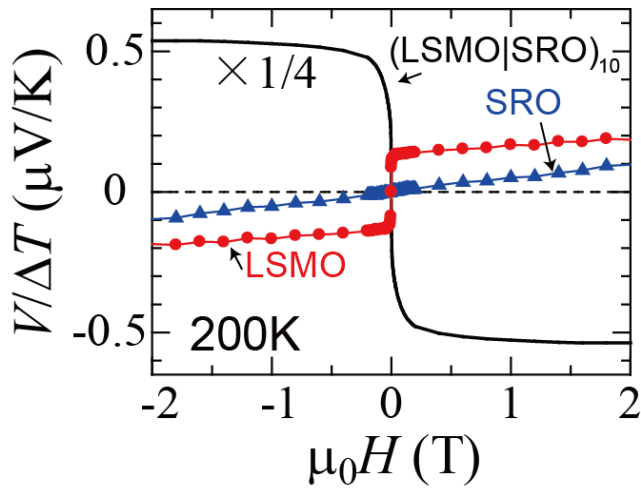}
\caption{Magnetic field ($H$) dependence of the transverse thermoelectric voltage divided by temperature difference ($V/\Delta T$) at $200$ K for a $50$-nm-thick La$_{0.67}$Sr$_{0.33}$MnO$_{3}$ (LSMO) film, a $50$-nm-thick SrRuO$_{3}$ (SRO) film, and a La$_{0.67}$Sr$_{0.33}$MnO$_{3}$$\mid$SrRuO$_{3}$ [(LSMO$\mid$SRO)$_{10}$] superlattice film.  }
\label{fig3}
\end{center}
\end{figure}

Figure \ref{fig2}(b) shows magnetization curves for the superlattice measured at several temperatures. The magnetization increases rapidly in a low-$H$ region and tends to saturate above $\sim 0.2$ T. Since La$_{0.67}$Sr$_{0.33}$MnO$_{3}$ has a small magnetic anisotropy and a low coercive field, magnetic hysteresis is hardly observed at $300$ K or $200$ K. On the other hand, SrRuO$_{3}$ possesses a strong uniaxial anisotropy and a large coercive field ($\geq 0.1$ T). In the $T$ range below $105$ K where the SrRuO$_{3}$ layers exhibit ferromagnetism, clear magnetic hysteresis of $M$ is observed.  
\par

We have measured transverse thermoelectric effects for the La$_{0.67}$Sr$_{0.33}$MnO$_{3}$$\mid$SrRuO$_{3}$ superlattice, as illustrated in Fig. \ref{fig2}(c). The sample was sandwiched by two AlN blocks; on the top AlN block, a $100$-${\rm \Omega}$-heater was attached to apply a temperature gradient to the superlattice, while the bottom block was kept at the system temperature. The temperature difference ($\Delta T$) which arises between the two AlN blocks was measured using type-E thermocouples. The magnitude of $\Delta T$ is $0.5$-$1$ K at each temperature. The thermoelectric voltage along the Hall direction induced by charge and spin transports from the hot to cold reservoir across the superlattice was measured between both ends of the film plane, as illustrated in Fig. \ref{fig2}(c). The length between the voltage electrodes and sample width are about $6$ mm and $2.5$ mm, respectively.
\par

Figure \ref{fig2}(d) presents the $H$ dependence of transverse thermoelectric voltage signal divided by the temperature difference, $V/\Delta T$, at $50$, $100$, $200$, and $300$ K. Clear voltage signal which is proportional to the magnetization is observed below $200$ K, while it is very small at $300$ K. The magnitude of $V/\Delta T$ increases with decreasing $T$ down to $100$ K, which is consistent with the $T$ variation of $M$ [Fig. \ref{fig2}(b)]. The sign of $V$ is negative for a positive magnetic field.   
\par

We found that the sign of $V/\Delta T$ for the superlattices is opposite to those for La$_{0.67}$Sr$_{0.33}$MnO$_{3}$ and SrRuO$_{3}$ single layers at the same temperature ($200$ K). Figure \ref{fig3} shows the magnetic-field dependence of Nernst voltage measured in the same experimental setup at $200$ K for a $50$-nm-thick La$_{0.67}$Sr$_{0.33}$MnO$_{3}$ film and a $50$-nm-thick SrRuO$_{3}$ film. At $200$ K, SrRuO$_{3}$ is paramagnetic and shows no ANE, while ferromagnetic La$_{0.67}$Sr$_{0.33}$MnO$_{3}$ shows an ANE. The magnitude of $V/\Delta T$ for the La$_{0.67}$Sr$_{0.33}$MnO$_{3}$ film is, however, much smaller than that for the La$_{0.67}$Sr$_{0.33}$MnO$_{3}$$\mid$SrRuO$_{3}$ superlattice (Fig. \ref{fig3}) in spite of their similar resistivity values ($\sim 0.5$ m${\rm \Omega cm}$). Furthermore, the sign of $V/\Delta T$ for the La$_{0.67}$Sr$_{0.33}$MnO$_{3}$ film is opposite to that for the La$_{0.67}$Sr$_{0.33}$MnO$_{3}$$\mid$SrRuO$_{3}$ supelattice. Hence, the transverse thermoelectric effect in the La$_{0.67}$Sr$_{0.33}$MnO$_{3}$$\mid$SrRuO$_{3}$ superlattices is not explained only by the ANE in La$_{0.67}$Sr$_{0.33}$MnO$_{3}$ or SrRuO$_{3}$, unless a drastic change in electronic structure around the Fermi energy, {\it e.g.} resonant states \cite{heremans}, is induced by charge transfer or strain effects at the interfaces. Generation of additional voltages due to spin-current generation effects, {\it e.g.} the SSE, may be important in the transverse thermoelectric voltage in the La$_{0.67}$Sr$_{0.33}$MnO$_{3}$$\mid$SrRuO$_{3}$ superlattices; in fact, very large ANE and SSE signals have been observed in other magnetic superlattices very recently \cite{raphael, KDLee}. 
\par

\begin{figure}[t]
\begin{center}
\includegraphics[width=8.5cm]{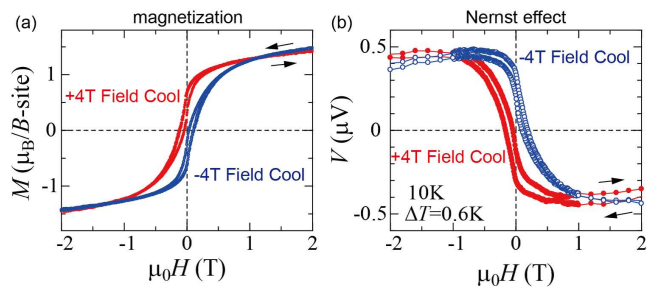}
\caption{Hysteresis loops at $10$ K for (a) magnetization ($M$) and (b) transverse thermoelectric voltage ($V$). The measurements were performed after the sample was cooled from $150$ K in an in-plane magnetic field of $+4$ T (red online) or after cooling from $150$ K in $-4$ T (blue online). The dotted lines are guides for the eyes.  } 
\label{fig4}
\end{center}
\end{figure}

Hysteresis loops of magnetization ($M$) and thermoelectric voltage along the Hall direction ($V$) at $10$ K are shown in Fig. \ref{fig4}. When the sample was cooled from $150$ K in the $+4$ T magnetic-field before the measurements, the magnetization loop is shifted in the negative direction along the field axis, as shown in Fig. \ref{fig4}(a). The direction of the shift is opposite after cooling from $150$ K in $-4$ T [Fig. \ref{fig4}(a)]. The magnitude of the shifts is about $0.1$ T. These results are direct evidence of the exchange bias effect. Because of the strong interfacial antiferromagnetic coupling, the magnetization of the SrRuO$_{3}$ layers are pinned in the opposite direction to the applied magnetic field in the field-cooling processes, which causes the negative exchange bias effect. 
Corresponding to the exchange bias observed in the magnetization curves at $10$ K, hysteresis loops of transverse thermoelectric voltages are also shifted from the zero field depending on the field-cooled processes, as shown in Fig. \ref{fig4}(b). The direction and magnitude of the shifts are consistent with those in magnetization curves [Fig. \ref{fig4}(a)]. Hence, the negative exchange bias effect is clearly observed also in the transverse thermoelectric effect driven by a cross-plane heat current. 
\par

In summary, we have studied the magnetothermoelectric effect for La$_{0.67}$Sr$_{0.33}$MnO$_{3}$$\mid$SrRuO$_{3}$ superlattices, in which ferromagnetic La$_{0.67}$Sr$_{0.33}$MnO$_{3}$ and SrRuO$_{3}$ couple antiferromagnetically below $105$ K. Clear voltage signals proportional to magnetization curves were observed below $300$ K. The sign of the transverse thermoelectric voltage for the superlattice is opposite to that for La$_{0.67}$Sr$_{0.33}$MnO$_{3}$ and SrRuO$_{3}$ films at $200$ K. Modulation of the ANE by a dramatic electronic reconstruction nucleated at the interfaces or generation of spin-current driven voltage in addition to the ANE may be important in the magnetothermoelectric voltage in the superlattices. At $10$ K, depending on magnetic-field directions in field-cooled processes, the magnetothermoelectric voltage loops are shifted from the zero field due to the exchange bias effect in the superlattices.  
\par

We thank K. Uchida for fruitful discussions and  S. Ito from Analytical Research Core for Advanced Materials, Institute for Materials Research, Tohoku University, for performing transmission electron microscopy on our samples. T.K. is supported by the Japan Society for the Promotion of Science (JSPS) through a research fellowship for young scientists. This work was supported by CREST ``Creation of Nanosystems with Novel Functions through Process Integration" from JST, Japan, and Grants-in-Aid for Scientific Research on Innovative Areas gNano Spin Conversion Scienceh (No. 26103005), Challenging Exploratory Research (No. 26610091) and Scientific Research (A) (No. 24244051) from MEXT, Japan.






\begin{thebibliography}{99}
\bibitem{uchida-nature}K. Uchida, S. Takahashi, K. Harii, J. Ieda, W. Koshibae, K. Ando, S. Maekawa, and E. Saitoh, Nature {\bf 455}, 778-781 (2008). 
\bibitem{kirihara}A. Kirihara, K. Uchida, Y. Kajiwara, M. Ishida, Y. Nakamura, T. Manako, E. Saitoh, and S. Yorozu, Nature Mater. {\bf 11}, 686-689 (2012).
\bibitem{bauer}G.E.W. Bauer, E. Saitoh, and B.J. van Wees, Nature Mater. {\bf 11}, 391-399 (2012).
\bibitem{heremans-review}S.R. Boona, R.C. Myers, and J.P. Heremans, Energy Environ. Sci. {\bf 7}, 885-910 (2014).
\bibitem{ong}W-L. Lee, S. Watauchi, V.L. Miller, R.J. Cava, and N.P. Ong, Phys. Rev. Lett. {\bf 93}, 226601 (2004).
\bibitem{xiao}D. Xiao, Y. Yao. Z. Fang, and Q. Niu, Phys. Rev. Lett. {\bf 97}, 026603 (2006).
\bibitem{chiba}Y. Pu, D. Chiba, F. Matsukura, H. Ohno, and J. Shi, Phys. Rev. Lett. {\bf 101}, 117208 (2008).
\bibitem{onoda}S. Onoda, N. Sugimoto, and N. Nagaosa, Phys. Rev. {\bf B} {\bf 77}, 165103 (2008).
\bibitem{hanasaki}N. Hanasaki, K. Sano, Y. Onose, T. Ohtsuka, S. Iguchi, I. K\'ezsm\'arki, S. Miyasaka, S. Onoda, N. Nagaosa, and Y. Tokura, Phys. Rev. Lett. {\bf 100}, 106601 (2008).
\bibitem{shiomi-Nernst}Y. Shiomi, N. Kanazawa, K. Shibata, Y. Onose, and Y. Tokura, Phys. Rev. B {\bf 88}, 064409 (2013).
\bibitem{tokura}M. Imada, A. Fujimori, and Y. Tokura, Rev. Mod. Phys. {\bf 70}, 1039 (1998).
\bibitem{hwang}H.Y. Hwang, Y. Iwasa, M. Kawasaki, B. Keimer, N. Nagaosa, and Y. Tokura, Nature Mater. {\bf 11}, 103-113 (2012).
\bibitem{heinze}S. Heinze, H.-U. Habermeier, G. Cristiani, S.B. Canosa, M.L. Tacon, and B. Keimer, Appl. Phys. Lett. {\bf 101}, 131603 (2012).
\bibitem{ziese}M. Ziese, I. Vrejoiu, E. Pippel, P. Esquinazi, D. Hesse, C. Etz, J. Henk, A. Ernst, I.V. Maznichenko, W. Hergert, and I. Mertig, Phys. Rev. Lett. {\bf 104}, 167203 (2010).
\bibitem{lee}Y. Lee, B. Caes, and B. N. Harmon, J. Alloys Compd. {\bf 450}, 1 (2008).
\bibitem{ke}X. Ke, M.S. Rzchowski, L.J. Belenky, and C.B. Eom, Appl. Phys. Lett. {\bf 84}, 5458 (2004).
\bibitem{qzhang}Q. Zhang, S. Thota, F. Guillou, P. Padhan, V. Hardy, A. Wahl, and W. Prellier, J. Phys.: Condens. Matter {\bf 23}, 052201 (2011).
\bibitem{padhan}P. Padhan and W. Prellier, Appl. Phys. Lett. {\bf 99}, 263108 (2011).
\bibitem{kim}J.-H. Kim, I. Vrejoiu, Y. Khaydukov, T. Keller, J. Stahn, A. R\"uhm, D. K. Satapathy, V. Hinkov, and B. Keimer, Phys. Rev. {\bf B} {\bf 86}, 180402(R) (2012).
\bibitem{ke2}X. Ke, L. J. Belenky, V. Lauter, H. Ambaye, C.W. Bark, C. B. Eom, and M.S. Rzchowski, Phys. Rev. Lett. {\bf 110}, 237201 (2013).
\bibitem{zieseAPL}M. Ziese, I. Vrejoiu, and D. Hesse, Appl. Phys. Lett. {\bf 97}, 052504 (2010).
\bibitem{zieseReview}M. Ziese and I. Vrejoiu, Phys. Status. Solidi RRL {\bf 7}, 243-257 (2013). 
\bibitem{kuchida}K. Uchida, H. Adachi, T. Ota, H. Nakayama, S. Maekawa, and E. Saitoh, Appl. Phys. Lett. {\bf 97}, 172505 (2010).
\bibitem{ziese2}M. Ziese, F. Bern, A. Setzer, E. Pippel, D. Hesse, and I. Vrejoiu, Eur. Phys. J. B {\bf 86}, 42 (2013).
\bibitem{heremans}J.P. Heremans, V. Jovovic, E.S. Toberer, A. Saramat, K. Kurosaki, A. Charoenphakdee, S. Yamanaka, and G.J. Snyder, Science {\bf 321}, 554-557 (2008).
\bibitem{raphael}R. Ramos, T. Kikkawa, M. H. Aguirre, I. Lucas, A. Anadon, T. Oyake, K. Uchida, H. Adachi, J. Shiomi, P. A. Algarabel, L. Morellon, S. Maekawa, E. Saitoh, and M.R. Ibarra, arXiv:1503.05594 (2015).
\bibitem{KDLee}K.-D. Lee, D.-J. Kim, H.Y. Lee, S.-H. Kim, J.-H. Lee, K.-M. Lee, J.-R. Jeong, K.-S. Lee, H.-S. Song, J.-W. Sohn, S.-C. Shin, and B.-G. Park, arXiv:1504.00642 (2015).
\end{thebibliography}
\end{document}